\documentclass[prd,onecolumn,aps,amsfonts,
floats,tightenlines,floatfix,superscriptaddress]{revtex4}
\usepackage[english]{babel}
\usepackage{epsfig}
\usepackage{psfrag}
\usepackage{graphicx}
\usepackage{amsmath}
\usepackage{amssymb}
\usepackage{dsfont}
\usepackage[dvips]{color}
\newcommand{\intq}{\int\!\!\frac{d^4q}{(2 \pi)^4}}

\newcommand{\pslash}{p\!\!\!/}

\newcommand{\phslash}{\hat{p}\!\!\!/\,}

\newcommand{\wpslash}{\omega_{p}\!\!\!\!\!\!/\,\,\,}

\begin{document}

\date{\today}

\title{Neutrality of the color-flavor--locked 
phase in a Dyson-Schwinger approach}

\author{D.~Nickel}
\affiliation{Institute for Nuclear Physics, Technical University Darmstadt, 
  Schlo{\ss}gartenstra{\ss}e 9, D-64289 Darmstadt, Germany}
\author{R.~Alkofer}
\affiliation{Institute of Physics, University of Graz,
  Universit{\"a}tsplatz 5, A-8010 Graz, Austria}
\author{J.~Wambach}
\affiliation{Institute for Nuclear Physics, Technical University Darmstadt, 
  Schlo{\ss}gartenstra{\ss}e 9, D-64289 Darmstadt, Germany}
\affiliation{Gesellschaft f{\"u}r Schwerionenforschung mbH, Planckstra{\ss}e
  1, D-64291 Darmstadt, Germany}

\begin{abstract}
The role of neutrality constraints for the phase structure of QCD at
non-vanishing chemical potentials is studied within a self-consistent
truncation scheme for the Dyson-Schwinger equation of the quark propagator in
Landau gauge. We find the (approximate) color-flavor--locked phase to be
energetically
preferred at all potentially relevant densities  and for physical values of the
quark masses. We furthermore observe the impossibility to define this phase by
residual global symmetries and discuss the role of chemical potentials.
\end{abstract}

\maketitle
\section{Introduction}
\label{sect:neutralCFL}
The phase diagram of quantum chromodynamics (QCD) is expected to contain 
color-superconducting regimes at sufficiently high densities and low 
temperatures~\cite{Rajagopal:2000wf,Alford:2001dt,Schafer:2003vz,Rischke:2003mt,Buballa:2003qv,Shovkovy:2004me,Alford:2007xm}.
Considering three (degenerate) quark flavors, the color-flavor locked (CFL)
phase~\cite{Alford:1998mk} is the ground-state for vanishing
temperatures and asymptotically large 
densities~\cite{Schafer:1999fe,Shovkovy:1999mr}.
At densities of potential relevance for compact stellar objects it is,
however, not obvious how the splitting of Fermi surfaces due to finite quark
masses and neutrality constraints is influencing the ground-state~\cite{Alford:1999pa,Alford:2002kj,Steiner:2002gx}.
In Ref.~\cite{Nickel:2006kc} we have elaborated the role of the renormalized 
strange-quark  mass when addressing this question. Including medium 
modifications of the effective quark interaction decreases the dynamically 
generated contribution to the strange-quark mass. Therefore, for the physical
value of the strange-quark mass the CFL phase has been found to be the
ground-state at all potentially relevant densities.

In this paper we will report on the results when incorporating neutrality 
conditions.
As the CFL phase is neutral in the chiral limit due to the residual symmetry,
we do not expect it to be influenced strongly. In contrast, the 2SC phase for
two flavors is far from being electrically neutral. Therefore, as known from
investigations in Nambu--Jona-Lasinio (NJL) type
models~\cite{Ruster:2005jc,Steiner:2002gx,Ruster:2004eg,Abuki:2005ms}, the
region of the 2SC phase is strongly affected by neutrality constraints.
When varying the strange-quark mass the corresponding phenomenology can be,
in a strongly simplified manner, sketched as follows: 
In the chiral limit, all quarks have the same Fermi
momentum, and due to Luttinger's theorem, also have the same density. 
Increasing the strange-quark mass the Fermi momentum of the strange
quarks is lowered according to $p_{F,s}\approx \mu-{M_{s}^{2}}/{2\mu}$. 
To first approximation the positive electrical charge of the up-quarks is 
then compensated by additional down-quarks with Fermi momentum
$p_{F,d}\approx \mu+{M_{s}^{2}}/{2\mu}$.
This mechanism is related to a partial symmetry breaking of the residual
symmetry in the CFL phase after the formation of Cooper pairs.
Eventually, in some channels ungapped quasiparticle
excitations appear, and accordingly these phases are
referred to as the gapless CFL (gCFL)
phase~\cite{Shovkovy:2003uu,Huang:2003xd,Alford:2003fq,Alford:2004hz}.
At some stage pairing of down- and strange-quarks is no longer favorable, and
all remaining Cooper pairs involve an up-quark, which is characterizing the
uSC phase.
After this, the pairing of up- and strange-quarks might break, and one
potentially may find in an intermediate regime the 2SC phase.
Finally, all Fermi surfaces are far separated and the ground-state is
no longer a spin-zero superfluid.
Ungapped uSC and 2SC phases have not been observed in our
calculations.
It is worth mentioning that gapless phases are expected to be unstable, as some
Meissner masses become
imaginary~\cite{Huang:2004bg,Huang:2004am,Casalbuoni:2004tb}. This
finding led to a broad discussion of the appearance of inhomogeneous
phases~\cite{Schafer:2005ym,Rajagopal:2006ig,Gorbar:2005rx,Hong:2005jv}.
Due to the employed truncations this will not be addressed in this
work.
As suggested by our previous work~\cite{Nickel:2006kc} we find however those
scenarios to be irrelevant for realistic quark masses with gaps as large as we
obtain.

This paper is organized as follows:
In section~\ref{sect:setup} we briefly present the theoretical
framework which involves a fully self-consistent treatment of the 
quark propagator in the CFL-like phase including neutrality constraints.
The role of chemical potentials especially in the CFL phase will be discussed
in section~\ref{sect:electronsinCFL}.
In section~\ref{sect:results} we discuss the numerical results for the
CFL-like phase at non-vanishing strange-quark masses.
Finally, we summarize and conclude in section~\ref{sect:conclusions}.
The color-flavor basis is given in an Appendix.

\section{The Landau-gauge quark propagator at non-vanishing chemical
potential}
\label{sect:setup}
\subsection{The truncated quark Dyson-Schwinger equation
 at non-vanishing chemical potential}
In Refs.~\cite{Nickel:2006vf,Nickel:2006kc} a closed truncated Dyson-Schwinger
equation (DSE)
for the Landau-gauge  quark propagator at non-vanishing chemical
potential has been solved in the chiral limit and for finite
strange-quark masses, respectively. In the present work we follow the same
scheme and notations. In the Nambu-Gor'kov basis (see
Ref.~\cite{Rischke:2003mt} for a review) the normal
quark DSE is coupled to an equation for the gap functions. To describe
their structure we define
\begin{eqnarray}
\mathcal{S}_{0}(p)&=&
\left(
  \begin{array}{cc}
    S_{0}^{+}(p) & 0\\
    0 & ~\hspace{4mm} S_{0}^{-}(p) = -CS_{0}^{+}(-p)^{T}C
  \end{array}
\right),\\
\mathcal{S}(p)&=&
\left(
  \begin{array}{cc}
    S^{+}(p)& T^{-}(p) = \gamma_{4}T^{+}(p)^{\dagger}\gamma_{4}\\
    T^{+}(p)& ~\hspace{4mm} S^{-}(p) = -CS^{+}(-p)^{T}C
  \end{array}
\right),\\
\Sigma(p)&=&
\left(
  \begin{array}{cc}
    \Sigma^{+}(p)&~\hspace{5mm} \Phi^{-}(p)\\
    \Phi^{+}(p)&~\hspace{5mm} \Sigma^{-}(p)
  \end{array}
\right),
\end{eqnarray}
with
$\mathcal{S}_{0}$ being the `bare' and $\mathcal{S}$ the full Nambu-Gor'kov
propagator. The Nambu-Gor'kov self-energy is denoted by $\Sigma$.
As the Euclidean action is real we furthermore have
to impose $\Phi^{-}(p)   = \gamma_{4}\Phi^{+}(p)^{\dagger}  \gamma_{4}$.
The inverse `bare' quark propagator in the presence of a
static, isotropic and homogeneous `external'
gluon field with non-vanishing time component $A_{4}$ is
given by
\begin{eqnarray}
 \mathcal{S}_{0}^{-1}(p) &=&
 \left(
   \begin{array}{cc}
     -i \vec{p}\cdot\vec{\gamma}-i(p_{4}+i\bar{\mu}+\frac{Z_{1F}}{Z_{2}}gA_{4})\gamma_{4} +m& 0\\
     0 & -i \vec{p}\cdot\vec{\gamma}-i(p_{4}-i\bar{\mu}-\frac{Z_{1F}}{Z_{2}}gA_{4}^{T})\gamma_{4}+m\\
   \end{array}
 \right)
 \,.
\end{eqnarray}
Here $m$ is the mass matrix and $\bar{\mu}$ the matrix implementing chemical
potentials associated to global symmetries. The
parameterization and determination of $\bar{\mu}$ and $A_{4}$ will be
discussed below.
The renormalization constants $Z_{2}$ and $Z_{1F}$ also appear in 
the DSE for the quark propagator
\begin{eqnarray}
\label{qDSE}
  \mathcal{S}^{-1}(p) &=& Z_{2}\mathcal{S}_{0}^{-1}(p)+ Z_{1F}\Sigma(p) 
\end{eqnarray}
which explicitly reads
\begin{eqnarray}
  \label{fullT}
  T^{\pm} &=&
  -Z_{1F}\left(Z_{2}{S^{\mp}_{0}}^{-1}+Z_{1F}\Sigma^{\mp}\right)^{-1}
  \Phi^{\pm}S^{\pm},\\
  \label{fullS}
  {S^{\pm}}^{-1} &=& \hphantom{-}Z_{2}{S^{\pm}_{0}}^{-1}+Z_{1F}\Sigma^{\pm}-
  Z_{1F}^{2}\Phi^{\mp}\left(Z_{2}{S^{\mp}_{0}}^{-1}+Z_{1F}\Sigma^{\mp}
  \right)^{-1}\Phi^{\pm}.
\end{eqnarray}
Employing the truncation of Ref.~\cite{Nickel:2006vf}, the equations for the 
self-energy and the gap function then result in
\begin{eqnarray}
  \Sigma^{+}(p) &=& \hphantom{-}\frac{Z_{2}^{2}}{Z_{1F}} \pi \intq
  \gamma_{\mu} \lambda_{a} S^{+}(q) \gamma_{\nu} \lambda_{a}
  \left(
    \frac{\alpha_{s}(k^{2}) P^{T}_{\mu\nu}}{k^{2}+G(k)}+
    \frac{\alpha_{s}(k^{2}) P^{L}_{\mu\nu}}{k^{2}+F(k)}
  \right),
  \\
  \Phi^{+}(p) &=& -\frac{Z_{2}^{2}}{Z_{1F}} \pi \intq
  \gamma_{\mu} \lambda_{a}^{T} T^{+}(q) \gamma_{\nu} \lambda_{a}
  \left(
    \frac{\alpha_{s}(k^{2}) P^{T}_{\mu\nu}}{k^{2}+G(k)}+
    \frac{\alpha_{s}(k^{2}) P^{L}_{\mu\nu}}{k^{2}+F(k)}
  \right),
\end{eqnarray} 
where $k=p-q$. Here projectors transverse and longitudinal to the medium have
been introduced. The functions $G$ and $F$ describe the corresponding
medium modifications of the gluon propagator, see Eq.~(21)
of Ref.~\cite{Nickel:2006vf}. They can be calculated once the coupling is given.
Therefore the only input for the quark DSE are the running coupling 
$\alpha_{s}(k^{2})$ and the renormalized current quark masses.

As the objective of the presented work is to show that the CFL phase
is the physical ground-state for realistic strange-quark masses, we
use the most conservative choice for $\alpha_{s}(k^{2})$ in our approach.
This is the running coupling $\alpha_{s}(k^{2})$ determined in DSE
studies of the Yang-Mills 
sector~\cite{Fischer:2002hn,Fischer:2002,Fischer:2003rp}. As is detailed
in Ref.~\cite{Fischer:2003rp} it underestimates chiral symmetry breaking
significantly in the Abelian approximation.
In Ref.~\cite{Nickel:2006vf} this coupling lead to the smallest
critical strange-quark masses.
However, we would like to emphasize that in contrast to NJL-tye
models, our approach is much less sensitive to the choice of the
coupling, an effect
which can be traced back to the inclusion of the medium polarization.

The renormalization constants are determined in the (chirally broken) vacuum.
Due to the vertex construction employed, the quark-gluon vertex renormalization
constant, $Z_{1F}$, cancels in the resulting renormalized equations.
For each flavor, we determine the quark wave-function renormalization
constant, $Z_{2}$, and the renormalization constant $Z_{m}$, relating the
unrenormalized quark mass $m_{0,q}(\Lambda^{2})$ at an ultraviolet 
cutoff $\Lambda$ to the renormalized mass $m_{q}(\nu)$ via
\begin{eqnarray}
  m_{0,q}(\Lambda^{2}) &=& Z_{m}(\nu^{2},\Lambda^{2})m_{q}(\nu),
\end{eqnarray}
by requiring
\begin{eqnarray}
  \left. S^{+}_{q}(p)\right|_{p^{2}=\nu^{2}} &=& -i\pslash+m_{q}(\nu)
\end{eqnarray}
at a renormalization scale $\nu$.
This corresponds to a momentum-subtraction ($MOM$) scheme, which results in
somewhat smaller numerical values for the quark current masses at a given
renormalization scale (usually taken to be $2\,\mathrm{GeV}$). We simply ignore
here the
difference between $\overline{MS}$ and $MOM$ masses because the effect is of
the order of ten percent (when calculated within perturbation theory) and is 
thus much smaller than the uncertainty in the value of the current masses.

It turns out that, as expected, the mass dependence of the quark wave function
renormalization constant, $Z_{2}$, is negligible as long as the renormalization
scale is much larger than the mass. Therefore, we simply drop this dependence
and $Z_{2}$ is determined once and for all in the chiral limit.  To keep the
number of parameters as small as possible, we set the up- and
down-quark masses to zero and vary the strange-quark current mass only.

In the following we will restrict to spatially isotropic phases. In order to solve the
DSE of the quark propagator it is advantageous to consider their color-flavor
structure first. To get a self-consistent solution we choose suitable sets of
matrices $\{P_{i}\}$ and $\{M_{i}\}$ in color-flavor space, such that
\begin{eqnarray}
  \Sigma^{+}(p)
  &=&
  \frac{Z_{2}}{Z_{1F}}
  \sum_{i} \Sigma^{+}_{i}(p) P_{i},\\
  \Phi^{+}(p)
  &=&
  \frac{Z_{2}}{Z_{1F}}
  \sum_{i} \phi^{+}_{i}(p) M_{i},
\end{eqnarray}
where we have introduced the renormalization-point independent component
functions $\Sigma^{+}_{i}(p)$ and $\phi^{+}_{i}(p)$, which are matrix-valued
in  Dirac space. Full self-consistency is guaranteed in case a basis of all
allowed matrices is considered. The dimensionality of this basis in a given phase
depends on the residual symmetry in color-flavor space. For the CFL phase
this will be detailed below, and the basis used is explicitely given
in the Appendix.

The Dirac structure of the self-energies in an even-parity phase can be
para\-meterized by~\cite{Pisarski:1999av}
\begin{eqnarray}
  \Sigma^{+}_{i}(p) &=&
  -i\phslash\,\Sigma^{+}_{A,i}(p)-i\wpslash\,\Sigma^{+}_{C,i}(p)
  +\Sigma^{+}_{B,i}(p)-i\gamma_{4}\phslash\,\Sigma^{+}_{D,i}(p), 
  \\
  \label{phi}
  \phi^{+}_{i}(p) &=& 
  \left(
    \gamma_{4}\phslash\,\phi^{+}_{A,i}(p)+\gamma_{4}\,\phi^{+}_{B,i}(p)
    +\phi^{+}_{C,i}(p)+\phslash\,\phi^{+}_{D,i}(p)
  \right)\gamma_{5},
\end{eqnarray}
where $\hat{p}=\vec{p}/\vert \vec{p}\vert$, $\phslash =\hat{p}\cdot\vec{\gamma}$, 
$\wpslash = \omega_{p}\gamma_{4}$ and $\omega_{p}=ip_{4}+\mu$.
Thus we finally need to solve a coupled set of integral equations
for the energy- and momentum-dependent functions $\Sigma^{+}_{ABCD,i}(p)$
and $\phi^{+}_{ABCD,i}(p)$. This is done numerically.

\subsection{Color neutrality}
As has been discussed
in~\cite{Gerhold:2003js,Dietrich:2003nu,Buballa:2005bv,Nickel:2006vf},
we need to allow for constant values of $A_{4}$, whose DSE reduces to
\begin{eqnarray}
  \label{eq:colorneutcondition}
  \rho^{a}(x)
  &\stackrel{\text{\tiny def}}{=}&
  Z_{2}\frac 1 2
  \int\!\!\frac{d^{3}p}{(2\pi)^{3}}
  \int\!\frac{dp_{4}}{2\pi}\,
  \mathrm{Tr}_{D,c,f,NG}
  \left(\mathcal{S}(p)\Gamma^{(0)a}_{NG4}\right)
  \nonumber\\
  &\stackrel{!}{=}&
  0\,.
\end{eqnarray}
As $A_{4}$ is anti-Hermitian within our conventions, we define by
\begin{align}
  \mu_{C}
  \quad\stackrel{\text{\tiny def}}{=}\quad
  \sum_{a}\mu_{a}\lambda^{a}
  \quad\stackrel{\text{\tiny def}}{=}\quad
  -i\frac{Z_{1F}}{Z_{2}}gA_{4}
\end{align}
the effective color chemical potentials. The color chemical potentials are
then adjusted to obtain color neutrality.

\subsection{Electrical neutrality and $\beta$-equilibrium}
Electrical neutrality within QCD alone is meaningless. We need to include the
electro-weak interaction and its particle content.
The conserved electrical charge is then adjusted by its Lagrange multiplier
$\mu_{Q}$ and the chemical potentials for the quarks, including the color
chemical potentials, are given by
\begin{eqnarray}
  \hat{\mu}_{ab,ij}
  &=& 
  \mu\,\delta_{ab}\delta_{ij}+
  \sum_{d}\mu_{d}\lambda^{d}_{ab}\delta_{ij}+
  \mu_{Q}\,\delta_{ab}Q_{ij}
  \,,
\end{eqnarray}
where $a,b=1,2,3$ denote color and $i,j=1,2,3$ flavor indices.
The charges of the quark flavors are encoded by the matrix
$Q=
\mathop{\mathrm{diag}}\nolimits_{f}
\left(\frac{2}{3},-\frac{1}{3},-\frac{1}{3}\right)
=\frac{1}{2}\tau^{3}+ \frac{1}{2\sqrt{3}}\tau^{8}$.
Thus the `bare' inverse quark propagator takes the form
\begin{eqnarray}
 \mathcal{S}_{0}^{-1}(p)
 &=&
 \left(
   \begin{array}{cc}
     -i \vec{p}_{i}\gamma_{i}-i(p_{4}+i\hat{\mu})\gamma_{4}+m & 0\\
     0 & -i \vec{p}_{i}\gamma_{i}-i(p_{4}-i\hat{\mu}^{T})\gamma_{4}+m\\
   \end{array}
 \right)
 \,.
\end{eqnarray}
Compared to the strong interaction, we can consider leptons as non-interacting
particles.
Only the electrons will be relevant, as the relevant charge chemical potentials
will not or only slightly exceed the mass of other charged leptons.
With charge $-1$, their charge chemical potential is
$\mu_{e}=-\mu_{Q}$. 
This also means
\begin{align}
  \mu_{d}
  \quad=\quad
  \mu_{s}
  \quad=\quad
  \mu_{u}+\mu_{e}
\end{align}
for each color, which is usually referred to as $\beta$-equilibrium.
We consider the electrons as massless and their electrical charge density
given by $\rho_{el}=\frac{1}{3\pi^{2}}\mu_{Q}^{3}$.
Electrical neutrality then enforces
\begin{eqnarray}
  \rho_{Q}(x)
  &=&
  Z_{2}
  \int\!\!\frac{d^{3}p}{(2\pi)^{3}}
  \int\!\frac{dp_{4}}{2\pi}\,
  \mathrm{Tr}_{D,c,f}
  \left(QS^{+}(p)\gamma_{4}\right)  
  +\frac{1}{3\pi^{2}}\mu_{Q}^{3}
  \nonumber\\
  &\stackrel{!}{=}&
  0
  \,.
\end{eqnarray}

\subsection{Parameterization of the CFL phase}
A finite strange-quark mass leads
to a partial symmetry breaking of the CFL symmetry in the chiral limit
through the mass matrix
$m=\frac 1 3 m_{s}\left(\mathds{1}-\sqrt{3}\tau^{8}\right)$.
If in addition $\mu_{Q}\neq 0$ due to the neutrality constraint, we can
directly conclude from its generator $Q=\frac{1}{2}\tau^{3}+
\frac{1}{2\sqrt{3}}\tau^{8}$ that
\begin{eqnarray}
  \label{eq:CFLsymmetries}
  SU_{c+V}(3)
  &\stackrel{m_{s}\neq 0}{\longrightarrow}&
  SU_{c+V}(2)\otimes U_{c+V}(1)
  \nonumber\\
  &\stackrel{\mu_{Q}\neq 0}{\longrightarrow}&
  U_{c+V}(1)\otimes U_{c+V}(1)
  \,,
\end{eqnarray}
where the residual symmetry is generated by $\tau_{3}-\lambda_{3}^{T}$ and
$\tau_{8}-\lambda_{8}^{T}$. Those form a Cartan subalgebra of the Lie algebra
of $SU_{c+V}(3)$,  i.e. a maximum set of commuting matrices.
The most general choice of a basis $\{P_{i}\}$ and
$\{M_{i}\}$, constructed as in Ref.~\cite{Nickel:2006mm} for a less
complicated ansatz,
is given in the Appendix. It consists of 15 matrices
for $\{P_{i}\}$ and $\{M_{i}\}$, respectively.

To preserve the residual $U_{c+V}(1)\otimes U_{c+V}(1)$ symmetry, we
are only allowed to vary the color chemical potentials $\mu_{3}$ and
$\mu_{8}$. The question is, whether this is enough to fulfill the
requirement $\rho^{a}(x)=0$ for $a=1,\dots,8$.
To clarify this, we define the color-charge density matrix $\hat{\rho}$ by
\begin{eqnarray}
  \hat{\rho}
  &=&
  \int\!\!\frac{d^{3}p}{(2\pi)^{3}}
  \int\!\frac{dp_{4}}{2\pi}\,
  \mathrm{Tr}_{D,f}
  \left(S^{+}(p)\gamma_{4}\right)
  \,.
\end{eqnarray}
This $3\times 3$ matrix is symmetric and can be interpreted as the matrix of
color-charges in the basis $\{u,d,s\}$. With its help the condition in
Eq.~(\ref{eq:colorneutcondition}) is reduced to
\begin{eqnarray}
  \label{eq:colormatrixneut}
  \mathrm{Tr}_{c}\left(\hat{\rho}\lambda^{a}\right)
  &=&
  0
  \,.
\end{eqnarray}
As the Gell-Mann matrices form an orthogonal basis, we conclude that we need to
adjust $\{\mu_{a}\}$ to get
\begin{eqnarray}
  \hat{\rho}
  &=&
  \frac{\rho}{3}\mathds{1}
  \,,
\end{eqnarray}
with $\rho$ then being the quark number density.

Varying $\mu_{3}$ and $\mu_{8}$, we will surely achieve the requirement in
Eq.~(\ref{eq:colormatrixneut}) if the matrix is diagonal.
This is the case for all NJL-type
investigations~\cite{Ruster:2005jc,Steiner:2002gx,Ruster:2004eg,Abuki:2005ms}.
However, this is a further truncation on the self-consistency
and enforced by hand.
In general the matrix $\hat{\rho}$ is only symmetric, if we also consider the
color-flavor structures $P_{4},\dots,P_{9}$. Therefore all $\mu_{a}$ for
$a=1,\dots,8$ need to be taken into account and the residual symmetry gets
completely broken.
The same arguments also hold if $\mu_{Q}=0$, but $\mu_{3}\neq 0$ or
$\mu_{8}\neq 0$.

We conclude, that a finite strange quark mass induces in general
backgrounds fields that break the CFL symmetry completely and the CFL
phase can no longer be defined by a residual symmetry.
It can only be defined by a continuity argument for the ground-state as a
function of $m_{s}$ that is CFL symmetric for $m_{s}=0$.

Instead of introducing further color chemical potentials aside from $\mu_{3}$
and $\mu_{8}$, we will estimate those as described below and choose $\mu_{3}$
and $\mu_{8}$ such that $\rho^{3}=0$ and $\rho^{8}=0$.
This approach is similar to the truncation being used in NJL-type
investigation and we could otherwise not constrain $\{P_{i}\}$ and
$\{M_{i}\}$.
The ansatz becomes self-consistent in the 2SC and uSC phase.

\subsection{Estimating $\mu_{a}$ not in the Cartan subalgebra}
Varying $\mu_{3}$ and $\mu_{8}$ only, we adjust the $3\times 3$-dimensional
real and positive charge-density matrix $\hat{\rho}$ to have equal diagonal
elements and we define
$\mathrm{Tr}\left(\hat{\rho}\right)=\frac{3}{\pi^2}\hat{p}_{F}^{3}$.
The off-diagonal elements of $\hat{\rho}$ are strongly suppressed.
After diagonalization we obtain
$\tilde{\rho}=
\mathop{\mathrm{diag}}\nolimits_{c}(\tilde{\rho}_{1},\tilde{\rho}_{2},\tilde{\rho}_{3})=
D^{\dagger}\hat{\rho}D$
and, approximating the system by a free gas, we estimate
$\tilde{\rho}_{i}=\frac{1}{\pi^2}\tilde{p}_{i}^3$.
For a free gas we would therefore need to vary $\tilde{\mu}_{3}$ and
$\tilde{\mu}_{8}$ in the new basis by
$\Delta\tilde{\mu}_{3}=\frac{1}{2}(\tilde{p}_{2}-\tilde{p}_{1})$ and
$\Delta\tilde{\mu}_{8}=
\frac{1}{2\sqrt{3}}(2\tilde{p}_{3}-\tilde{p}_{1}-\tilde{p}_{2})$
in order to obtain a neutral phase.
This will be used as an estimate of $\mu_{a}$ for $a\neq 3,8$
after the transformation into the old color basis.

The eigenvalues of $\hat{\rho}$ can be estimated in an expansion in
$\hat{\rho}_{ij}\ll \rho_{11}$ for $i\neq j$ and are given by
$\hat{\rho}_{11}$ and $\hat{\rho}_{11}\pm\delta\hat{\rho}$, with
$\delta\hat{\rho}=
\sqrt{\hat{\rho}_{12}\hat{\rho}_{21}+
\hat{\rho}_{23}\hat{\rho}_{32}+\hat{\rho}_{13}\hat{\rho}_{31}}$.
The corresponding Fermi momenta $\tilde{p}_{i}$ are approximately given by
$\hat{p}_{F}$ and
$\hat{p}_{F}(1\pm{\delta\hat{\rho}}/{3\hat{\rho}_{11}})$.
Therefore, after ordering the Fermi momenta, we have:
$\Delta\tilde{\mu}_{3}\approx{\hat{p}_{F}\delta\hat{\rho}}/{3\hat{\rho}_{ii}}$ and
$\Delta\tilde{\mu}_{8}\approx 0$. As for a free gas we also suspect the
chemical potentials to transform like
$\tilde{\mu}=D^{\dagger}\hat{\mu}D$ with $DD^{\dagger}=\mathds{1}$.
Thus
$\Vert\Delta\tilde{\mu}\Vert=
\Vert\tilde{\mu}-\mu\mathds{1}\Vert=
\Vert\hat{\mu}-\mu\mathds{1}\Vert=
\Vert\Delta\hat{\mu}\Vert$
for any matrix norm and we obtain
\begin{align}
  \mu_{a}
  \quad\lesssim\quad
  \Vert\Delta\tilde{\mu}\Vert
  \quad\approx\quad
  \frac{\delta\hat{\rho}}{3\hat{\rho}_{ii}}\,\hat{p}_{F}
  \,,\quad
  a\neq 3,8
  \,.
\end{align}
We will come back to this estimate when discussing our numerical results for
the chemical potentials.

\section{Electrons in the CFL phase and gapless pairing}
\label{sect:electronsinCFL}
\subsection{Are there electrons in the CFL phase?}
It has been argued that the color-neutral CFL phase is automatically
electrically neutral~\cite{Rajagopal:2000ff}. Therefore $\mu_{Q}=0$ and no
electrons are allowed in the phase, which would have important consequences.
The same result is also found in self-consistent NJL-type
investigations~\cite{Ruster:2005jc,Abuki:2005ms}. Since we find deviations, we
first describe the reasoning in those models:

Apart from the $SU_{c}(3)\otimes SU_{V}(3)$, we can also consider the
electrical $U_{Q}(1)$ generated by $Q$.
In the dynamical symmetry breakdown
$SU_{c}(3)\otimes SU_{V}(3)\rightarrow U_{c+V}(1)\otimes U_{c+V}(1)$ also
$U_{Q}(1)$ gets broken, however the phase is still symmetric under
the so-called $U_{\tilde{Q}}(1)$ symmetry that is generated by
\begin{eqnarray}
  \tilde{Q}
  &=&
  Q-\frac{1}{2}\lambda^{3}-\frac{1}{2\sqrt{3}}\lambda^{8}
  \nonumber\\
  &=&
  \frac{1}{2}\left(\tau^{3}-\lambda^{3 T}\right)
  +\frac{1}{2\sqrt{3}}\left(\tau^{8}-\lambda^{8T}\right)
  \,.
\end{eqnarray}
In the basis  $((r,u),(g,d),(b,s),(r,d),(g,u),(r,s),(b,u),(g,s),(b,d))$ used
in the Appendix, the
matrix $\tilde{Q}$ is diagonal and the quasiparticles carry the charges
$(0,0,0,-1,1,-1,1,0,0)$, respectively.
From this it has been concluded that $\rho_{\tilde{Q}}=0$ in the fully gapped
CFL phase, which then is a $\tilde{Q}$-insulator~\cite{Alford:2002kj}.
The quark contribution to the thermodynamical potential
$p_{q}[T,\mu,\mu_{3},\mu_{8},\mu_{Q}]$ would then be invariant under
\begin{eqnarray}
  p_{q}[T,\mu,\mu_{3},\mu_{8},\mu_{Q}]
  &=&
  p_{q}[
  T,\mu,\mu_{3}-\frac{1}{2}\mu_{\tilde{Q}},
  \mu_{8}-\frac{1}{2\sqrt{3}}\mu_{\tilde{Q}},\mu_{Q}+\mu_{\tilde{Q}}]
  \,.
\end{eqnarray}
Neglecting the leptons, we would therefore have degenerate ground-states
under variation of $\mu_{\tilde{Q}}$. Among those the real ground-state among
those is then chosen
by the minimum of the electronic (or leptonic) contribution to the
thermodynamic potential,
$p_{el}[T,\mu,\mu_{3},\mu_{8},\mu_{Q}]=\frac{1}{12\pi^{2}}\mu_{Q}^{4}$.
We would therefore conclude $\mu_{Q}=0$, which means that no electrons are
allowed in the system.

The whole argument is therefore based on the assumption $\rho_{\tilde{Q}}=0$.
Considering the form of the charge $\tilde{Q}$, we can concentrate on the
separate pairing of $\{(r,d),(g,u)\}$ and
$\{(r,s),(b,u)\}$, respectively, as only those carry non-vanishing
$\tilde{Q}$-charge (see Appendix).
The requirement $\rho_{\tilde{Q}}=0$ is therefore equivalent to the statement,
that two pairing fermion species with different chemical potentials have the
same density in a fully gapped phase. We will now show that this is true for
energy-independent gap functions, which are used in NJL-type models and
emphasize that energy-dependent gap functions, as in our framework, do not
require this.

For the case of two fermion species $a,b$ we neglect normal self-energies
and consider
\begin{eqnarray}
  \label{eq:twofermionS}
  \left(
    \begin{array}{cccc}
      S_{a}^{+} & & & T_{a}^{-}\\
      &S_{b}^{+}& T_{b}^{-} &\\
      &T_{a}^{+}& S_{a}^{-}&\\
      T_{b}^{+}& & & S_{b}^{-}
    \end{array}
  \right)^{-1}
  &=&
  \left(
    \begin{array}{cccc}
      -i\pslash+\mu_{a}\gamma_{4}& & &-\gamma_{5}\Delta^{*}\\
      &-i\pslash+\mu_{b}\gamma_{4}& -\gamma_{5}\Delta^{*}&\\
      &  \gamma_{5}\Delta&-i\pslash-\mu_{a}\gamma_{4}&\\
       \gamma_{5}\Delta& & &-i\pslash-\mu_{b}\gamma_{4}
    \end{array}
  \right)\,,
  \nonumber\\&&
\end{eqnarray}
giving
\begin{eqnarray}
  \left.S_{a/b}^{+}\right.^{-1}
  &=&
  -i\pslash+\mu_{a/b}\gamma_{4}
  -\vert\Delta\vert^{2}
  \frac{
    i\pslash+\mu_{b/a}\gamma_{4}
  }{
    (p_{4}-i\mu_{b/a})^{2}+\vec{p}^{2}
  }
  \,.
\end{eqnarray}
With the definition of $D_{a/b}=(p_{4}+i\mu_{a/b})^{2}+\vec{p}^{2}$ and
$\omega_{a/b}=p_{4}+i\mu_{a/b}$, the
density $\rho_{a/b}(x)$ of the fermions species $a$ and $b$ turns out to be
\begin{eqnarray}
  \label{eq:twofermionp4}
  \rho_{a/b}(x)
  &=&
  \int\!\!\frac{d^{3}p}{(2\pi)^{3}}
  \int\!\frac{dp_{4}}{2\pi}\,
  \mathrm{Tr}_{D}
  \left(S_{a/b}^{+}(p)\gamma_{4}\right)
  \nonumber\\
  &=&
  \int\!\!\frac{d^{3}p}{(2\pi)^{3}}
  \int\!\frac{dp_{4}}{2\pi}\,
  \frac{
    4iD^{*}_{b/a}
    \left(\omega_{a/b}D^{*}_{b/a}+\vert\Delta\vert^{2}\omega^{*}_{b/a}\right)
  }{
    \left(
      \omega_{a/b}D^{*}_{b/a}+
      \omega^{*}_{b/a}\vert\Delta\vert^{2}
    \right)^{2}
    +
    \vec{p}^{2}
    \left(
      D^{*}_{b/a}+\vert\Delta\vert^{2}
    \right)^{2}
  }
  \,.
\end{eqnarray}
For an energy independent gap function $\Delta$, we can perform the energy
integral and obtain for $\vert\mu_{a}-\mu_{b}\vert<2\vert\Delta\vert$
\begin{eqnarray}
  \label{eq:twofermionp3}
  \rho_{a/b}(x)
  &=&
  \int\!\!\frac{d^{3}p}{(2\pi)^{3}}
  \left(
    \frac{
      \left(\bar{\mu}-p\right)
    }{
      \sqrt{\left(p-\bar{\mu}\right)^{2}+\vert\Delta\vert^{2}}
    }
    +
    \frac{
      \left(\bar{\mu}+p\right)
    }{
      \sqrt{\left(p+\bar{\mu}\right)^{2}+\vert\Delta\vert^{2}}
    }
  \right)
  \,,
\end{eqnarray}
where $\bar{\mu}=\frac{1}{2}\left(\mu_{a}+\mu_{b}\right)$. The momentum
integral is finite for a gap function that vanishing sufficiently fast and
the result shows that for small differences in the chemical potential the
densities of both particle species are the same. This has also been found
in Ref.~\cite{Rajagopal:2000ff}.
For larger splittings, i.e.,
$\vert\mu_{a}-\mu_{b}\vert>2\vert\Delta\vert$, gapless modes emerge (which
will be discussed below).

As we have a non-trivial energy dependence of the gap functions the above
argument does no longer hold and we may find qualitative differences from
NJL-type investigations.
This may be directly verified by using an ansatz for the gap functions.
The presence of electrons is therefore also connected to a finite width in the
spectral function of the quasiparticles.
The latter is included in our investigations and is discussed in
Ref.~\cite{Nickel:2006mm} for the 2SC phase.

We would like to clarify that a finite width of the quasiparticles near
the Fermi surface may be an artefact of our current truncation, especially
since we have not implemented the Meissner effect and the interaction via
magnetic gluons is long-ranged.
A closer investigation should be subject of future work.

\subsection{Gapless pairing}
\label{subsect:gapless}
The discussion of the last subsection sheds also light on gapless pairing.
For illustration we will again use the simple parameterization in
Eq.~(\ref{eq:twofermionS}).
If the difference in the chemical potentials
$\delta\mu=(\mu_{a}-\mu_{b})/2$ exceeds the value $\vert\Delta\vert$, the gap
in
the excitation spectrum vanishes and we find gapless modes even for
$\Delta\neq 0$. Their dispersion relation are given by
$\det S^{+\,-1}_{a/b}=0$, which are the zeros of the denominator of the
integrand in Eq.~(\ref{eq:twofermionp4}).
For the CFL phase, gapless pairing might occur in some
channels~\cite{Alford:1999xc} and has been found as the preferred homogeneous
ground-state in NJL-type
investigations~\cite{Shovkovy:2003uu,Huang:2003xd,Alford:2003fq,Alford:2004hz}
due to neutrality constraints.
However, they are expected to be unstable as the Meissner masses become
imaginary~\cite{Huang:2004bg,Huang:2004am,Casalbuoni:2004tb}. This finding led
to a broad discussion of the appearance of inhomogeneous
phases~\cite{Schafer:2005ym,Rajagopal:2006ig,Gorbar:2005rx,Hong:2005jv}.
We will postpone this subject to future investigations and consider here
homogeneous phases only.
The problem of imaginary Meissner masses in the gap equation has not been
addressed yet and does not appear in our truncation either.

For $\vert\delta\mu\vert>\vert\Delta\vert$ gapless modes create a
breached pairing region~\cite{Gubankova:2003uj}. For the occupation numbers
implicitly given in Eq.~(\ref{eq:twofermionp3}) and choosing
$\mu_{a}<\mu_{b}$, we get
\begin{eqnarray}
  \label{eq:occnumbergapless}
  n_{a/b}(p)
  &=&
  \left\{
    \begin{array}{l}
      \displaystyle{\frac{
        \bar{\mu}+p\mp\sqrt{\left(p+\bar{\mu}\right)^{2}+\vert\Delta\vert^{2}}
      }{
        2\sqrt{\left(p+\bar{\mu}\right)^{2}+\vert\Delta\vert^{2}}
      }}
      \,,\quad \mathrm{for} \quad
      \bar{\mu}-2\sqrt{\delta\mu^{2}-\vert\Delta\vert^{2}}
        <p<
        \bar{\mu}+2\sqrt{\delta\mu^{2}-\vert\Delta\vert^{2}},
      \\
      \displaystyle{\frac{
        \left(\bar{\mu}-p\right)
      }{
        2\sqrt{\left(p-\bar{\mu}\right)^{2}+\vert\Delta\vert^{2}}
      }
      +
      \frac{
        \left(\bar{\mu}+p\right)
      }{
        2\sqrt{\left(p+\bar{\mu}\right)^{2}+\vert\Delta\vert^{2}}
      }}
      \,,\quad
      \mathrm{else}.
    \end{array}
  \right.
\end{eqnarray}
We see that in an interval of the Fermi momentum the occupation number is
almost vanishing for one species and is almost unity for the other.
The deviation of the occupation numbers from zero and unity in this interval
for our simple parameterization is solely coming from the anti-quasiparticles
and therefore of order ${\Delta^{2}}/{\mu^{2}}\ll 1$.
The physical interpretation is that the quasiparticles in this `breached
pairing region' do not pair.

\section{Numerical results}
\label{sect:results}
\begin{figure}
  {\includegraphics[height=6.cm]{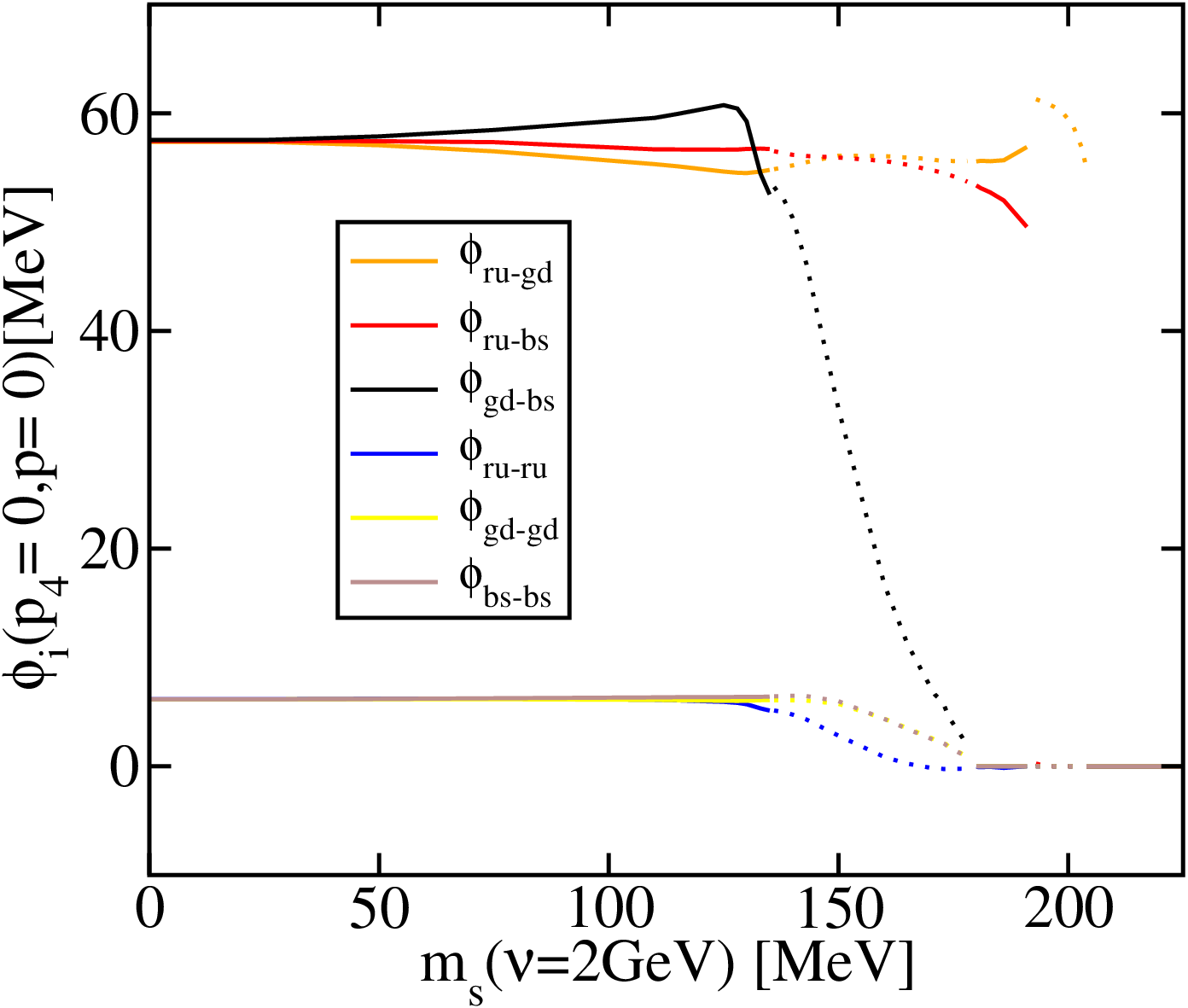}}
  \hspace{.8cm}
  {\includegraphics[height=6.cm]{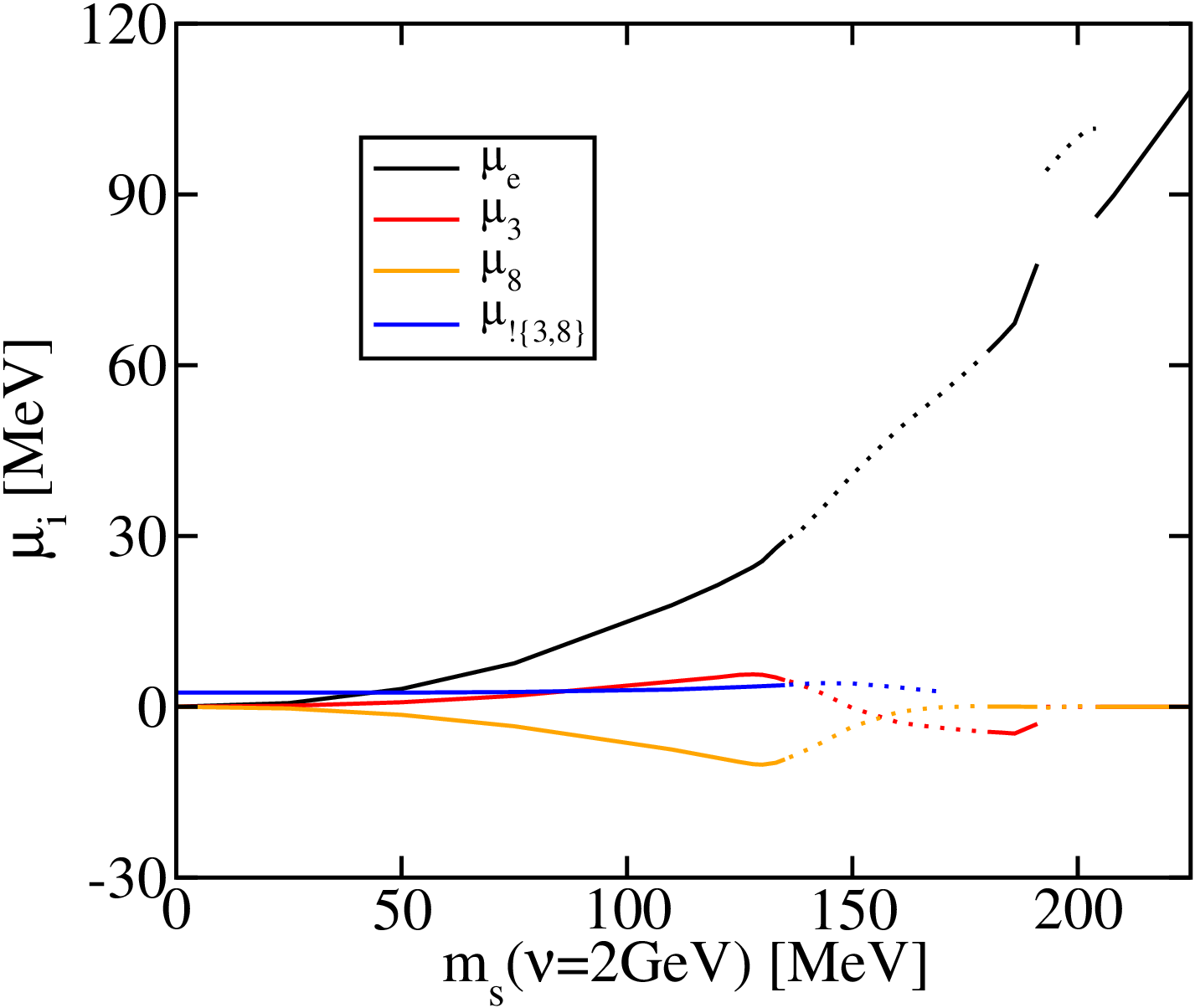}}
  \caption{Dependence of certain gap functions
  $\phi^{+}_{C,i}$ at the Fermi energy and for vanishing momentum (left) and of
  the (effective) chemical potentials (right) on the renormalized
  strange-quark mass. Both dependencies are determined at
  $\mu=400\,\mathrm{MeV}$.}
  \label{fig:gapmuneut}
\end{figure}
\begin{figure}
  {\includegraphics[height=6.cm]{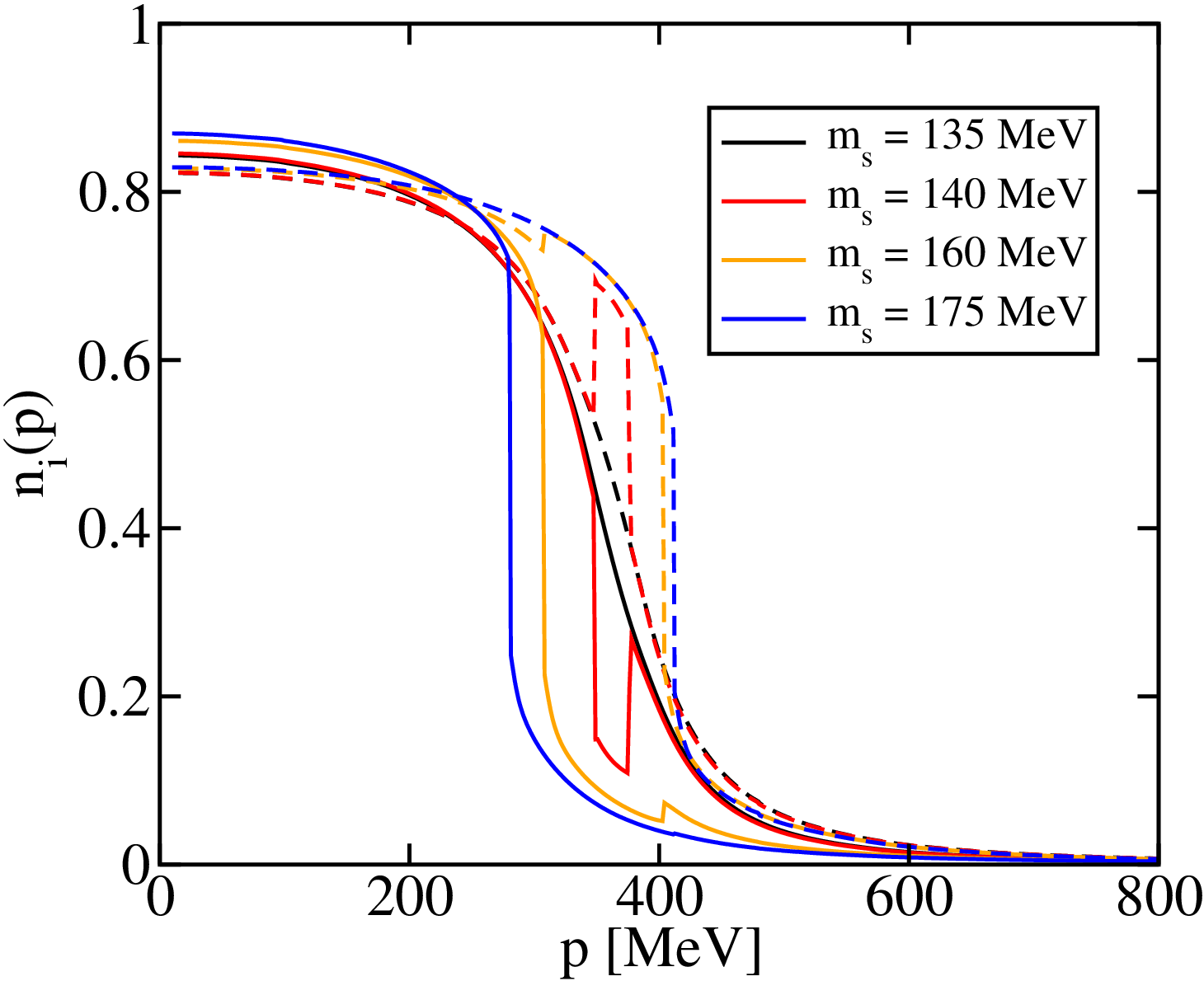}}
  \hspace{.8cm}
  {\includegraphics[height=6.cm]{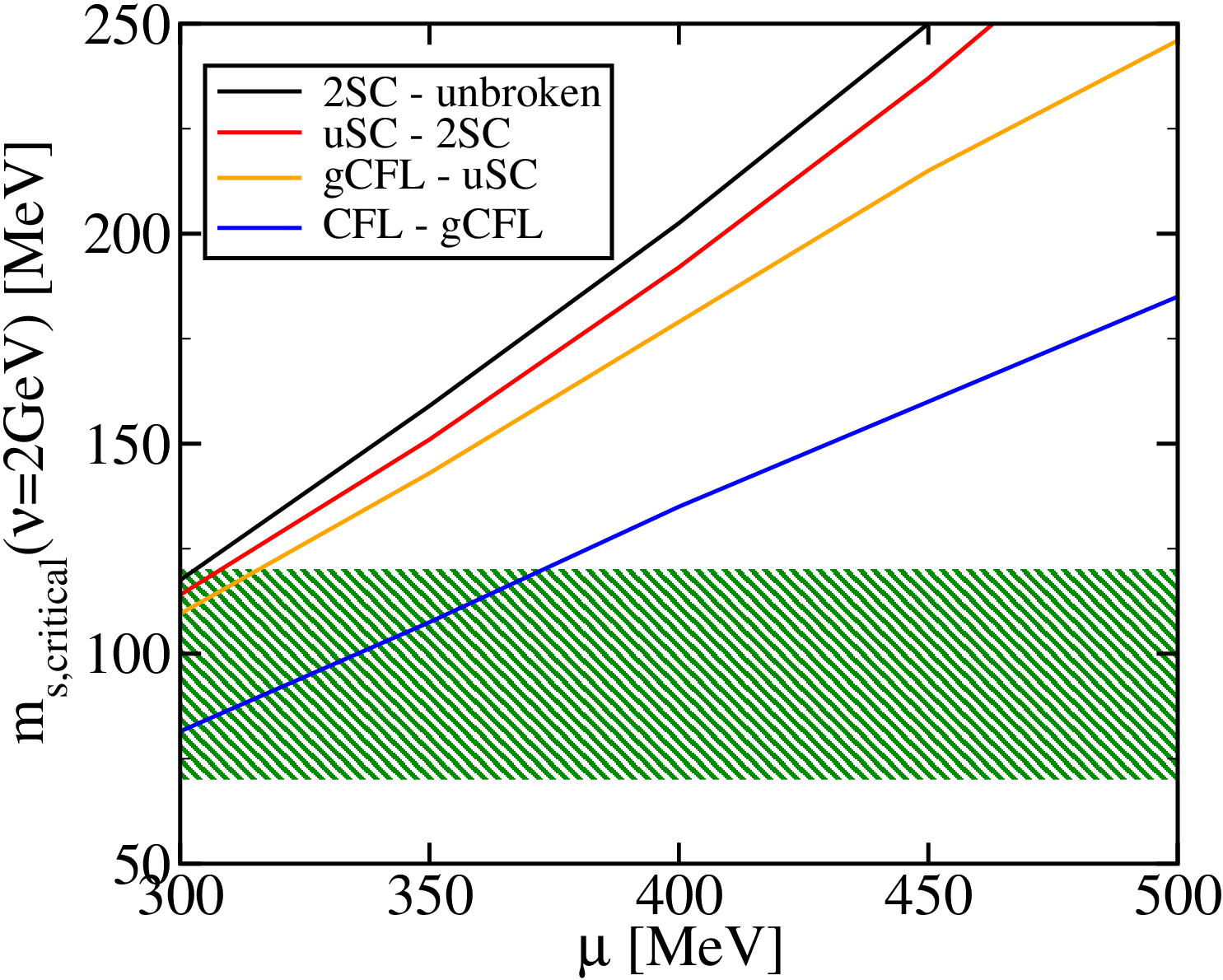}}
  \caption{Occupation numbers of green strange-quarks (solid) and blue
    down-quarks (dashed) for various values of the
    renormalized strange-quark mass at a chemical potential of
    $\mu=400\,\mathrm{MeV}$ (left) and the
    critical renormalized strange-quark masses as a function of the
    chemical potential together with the range stated by the particle data
    group~\cite{PDG06} (shaded area) (right).}
  \label{fig:np3mscritneut}
\end{figure}

We will now present numerical results for the neutral CFL phase. The
amount of data is overwhelming, as we calculate $4\times 15$ dressing functions
depending on energy and momentum for normal self-energy, gap function,
normal propagator and anomalous propagator, respectively.
However, we refer to Refs.~\cite{Nickel:2006vf,Nickel:2006kc} for
details and will only highlight new features coming from the neutrality constraints.

In Fig.~\ref{fig:gapmuneut} we present the dependence of the gap functions  on
the renormalized strange-quark mass taken at
the Fermi energy, i.e., for $p_{4}=0$, and for vanishing momentum,.
As the color-flavor structure of the pairing is quite involved, we refrain
from a determination of the Fermi momenta as we have done in 
Ref.~\cite{Nickel:2006kc}
and focus on the gap functions at vanishing  momentum. This is
reasonable as the gap functions $\phi^{+}_{C,i}$ for moderate chemical
potentials are almost constant below the Fermi energy (see
Refs.~\cite{Nickel:2006vf,Nickel:2006kc}).
The gap functions shown belong to the tensor structures $P_{1}$, $P_{2}$,
$P_{3}$, $P_{4}$, $P_{6}$ and $P_{7}$ given in the
Appendix and are labeled according to the quarks involved.
As can be seen, the gap functions are weakly dependent on the renormalized
strange-quark mass up to values of $\sim 135\,\mathrm{MeV}$. At this value, the
pairing of green strange-quarks with blue down-quarks becomes ungapped as
discussed in section~\ref{subsect:gapless}. This can also be observed in
the corresponding occupation numbers (see Fig.~\ref{fig:np3mscritneut}).
From the value $\sim 135\,\mathrm{MeV}$ 
onwards, the pairing gap of strange- and down-quarks `melts' and above
$\sim 180\,\mathrm{MeV}$ we are in the uSC phase. This phase exists up to a
value of $\sim 190\,\mathrm{MeV}$ and in the window of $\sim
190-200\,\mathrm{MeV}$ we find the 2SC phase. Above this value no
spin-zero pairing takes place anymore.

For numerical reasons we have not been able to compute the pressure difference
between the phases. However, the transitions of CFL to gCFL phase, and from 
there to the uSC phase, are likely to be at most of second order because the
self-energy and gap functions change continuously. In addition, if the intrinsic
symmetry is not altered, the transitions might even be crossovers. The
transitions of uSC to 2SC phase, and of the 2SC to the unpaired phase look like
first-order transitions. Here we also refer to our earlier observation, see
Ref.~\cite{Nickel:2006kc}, that in these phases it is much more difficult to
find an energetically disfavored solution. (As we solve the quark propagator
DSE, being the variation of the CJT action, by iteration in the
high-dimensional space of `discretized' dressing functions, the domain of
attraction for the global minimum is as usual strongly dominating.)  More
important, we want  emphasize that those transitions are not in the physically
relevant range of the strange-quark mass.

On the rhs of Fig.~\ref{fig:gapmuneut} we present the electron
chemical potential $\mu_{e}$, the color chemical potentials $\mu_{3}$ and
$\mu_{8}$ as well as an upper bound for the chemical potentials not belonging
to the Cartan subalgebra being labeled $\mu_{!\{3,8\}}$. It becomes obvious that electrical neutrality puts
the strongest constraint on the phase structure and $\mu_{e}$ even exceeds the
size of the largest gaps in the gCFL, uSC and 2SC phase. It is remarkable to
note that $\mu_{e}$ is non-vanishing in the gapped CFL phase as discussed in
section~\ref{sect:electronsinCFL}. This can also be seen in
Fig.~\ref{fig:np3mscritneut}, which shows (in this case for green strange- and
blue down-quarks) that the occupation numbers of pairing quarks in the gapped
phase need not be identical as concluded for energy independent gap functions
in Eq.~(\ref{eq:occnumbergapless}). The bound for the chemical potentials not
belonging to the Cartan subalgebra is of the order as the chemical $\mu_{3}$
and $\mu_{8}$. This shows that all color chemical potentials are much smaller
than the electron chemical potential $\mu_{e}$. Thus we estimate that the
treatment of $\mu_{3}$ and $\mu_{8}$ alone is not much better justified than
neglecting all color chemical potentials.

In Fig.~\ref{fig:np3mscritneut} the occupation numbers for the ungapped pairing
of green strange- and blue down-quarks at a chemical potential of
$\mu=400\,\mathrm{MeV}$ and for various renormalized strange-quark masses are
presented. Qualitatively those show a breached pairing region as found in
Eq.~(\ref{eq:occnumbergapless}) for a simple model study. Due to the
interaction, these occupation numbers in the breached pairing region are
neither almost vanishing nor close to unity, but more or less close to the
occupation numbers in the unpaired phase (see also Ref.~\cite{Nickel:2006vf}).

Last but not least, in Fig.~\ref{fig:np3mscritneut} we present the critical
values of the renormalized strange-quark mass, i.e., the values of the
renormalized strange-quark mass which separate the different phases, as a
function of the quark chemical potential. As compared to our analysis given in
Ref.~\cite{Nickel:2006kc} the transition from the CFL phase is slightly shifted
towards higher values of the strange-quark mass when imposing neutrality
constraints. This is indeed the anticipated result: The self-energy and gap
functions in the CFL phase are only weakly modified by enforcing neutrality
constraints, whereas these functions are strongly affected in the phases at
large strange-quark masses. In agreement with the general arguments given in
the introduction, towards larger strange-quark masses we first find the uSC
phase followed by the 2SC phase. Above some critical value of the renormalized
strange-quark mass none of the superfluid phases within the class of our
ans\"atze is energetically favored anymore. This is the case because the Fermi
surfaces of the different flavors are to far separated to allow for a spin-zero
pairing. In this region one expects pairing of quarks with the same flavor
only~\cite{Schafer:2000tw,Schmitt:2004et}, see also
Ref.~\cite{Marhauser:2006hy} for a treatment in the Dyson-Schwinger approach.
We also find a gCFL phase, which is supposed to be unstable (cf. the discussion
given in the introduction). For the physical strange-quark mass, given as a
range of values by the particle data group~\cite{PDG06}, this becomes only
relevant for the stated upper limit of the strange-quark mass and
quark chemical potentials $\lesssim 370\mathrm{MeV}$. For those values NJL-type
calculations~\cite{Ruster:2005jc,Steiner:2002gx,Ruster:2004eg,Abuki:2005ms}
typically find the chirally broken phase. (In order to achieve a conservative estimate
for color-flavor unlocking this does not yet take place in the presently used
truncation due to our approximation for the medium polarization.) In addition,
it is worth mentioning that solutions of the Bethe-Salpeter equation within the
DSE approach in the $MOM$ scheme favor small values for the physical strange
quark-current mass~\cite{Fischer:2005en}, a fact which is also related to the
difference in renormalization schemes. Taking into account all these facts we
conclude that among all possible color-superconducting phases only the CFL
phase is realized for physically relevant strange-quark masses and chemical
potentials.

\section{Conclusions}
\label{sect:conclusions}
As could have been anticipated from our previous investigation of color-flavor
unlocking, the CFL phase stays the ground-state at vanishing temperatures and 
above the critical chemical potential for the chiral phase  transition also
when imposing neutrality constraints. The underlying reason for this result,
which deviates from the one obtained from NJL-type models, is the inclusion
of medium polarization effects. The latter play an important role in presence of a
Fermi surface and lead to smaller dynamical mass generation due to damping and
screening~\cite{Nickel:2006kc}. As a consequence a stronger explicit symmetry
breaking by the bare strange-quark mass as present in nature  would be needed
to separate the Fermi surfaces and to `unlock' the CFL phase. To reinforce our
conclusions we have used a coupling, which is rather weak and not able to give
physical values for the chiral condensate and pion decay constant in the vacuum
and which thus should favor color-flavor unlocking.

Furthermore, it has been clarified that for a non-vanishing strange-quark mass 
the CFL phase cannot be defined by a residual global symmetry, and thus should better
be named color-flavor-locked-like phase. This can be traced back to the fact
that non-vanishing chemical potentials $\mu_{3}$, $\mu_{8}$ or $\mu_{Q}$  
induce static gluonic background fields which break the residual symmetry. 
This effect is, however, estimated to be small.

A surprising feature is the appearance of electrons in the CFL phase which has
been expected not to take place. Their presence is allowed due to the
energy-dependence of self-energy and gap functions which give the
quasiparticles a considerable width but may an artefact of the employed
truncation.

\section*{Acknowledgments}
We thank Michael Buballa and Krishna Rajagopal for helpful discussions and
comments. This work has been supported in part by the Helmholtz-University
Young Investigator Grant VH-NG-332. D.N. greatly appreciates the hospitality
at the CTP of MIT where this work has been finished.

\appendix*
\section{The color-flavor structure of the CFL phase}
For the neutral CFL phase with two degenerate quarks and non-vanishing
strange-quark mass, we generalize the ansatz of
Refs.~(\cite{Alford:1999pa,Nickel:2006kc}) by choosing the matrices
\begin{eqnarray}
  P_{i} &=&
  \left(
    \begin{array}{ccccccccc}
      \delta_{i1}&\delta_{i4}&\delta_{i6}&&&&&&\\
      \delta_{i5}&\delta_{i2}&\delta_{i7}&&&&&&\\
      \delta_{i8}&\delta_{i9}&\delta_{i3}&&&&&&\\
      &&&\delta_{i10}&&&&&\\
      &&&&\delta_{i11}&&&&\\
      &&&&&\delta_{i12}&&&\\
      &&&&&&\delta_{i13}&&\\
      &&&&&&&\delta_{i14}&\\
      &&&&&&&&\delta_{i15}\\
    \end{array}
  \right)
  \,,
\end{eqnarray}
\begin{eqnarray}
  M_{i} &=&
  \left(
    \begin{array}{ccccccccc}
      \delta_{i1}&\delta_{i4}&\delta_{i6}&&&&&&\\
      \delta_{i5}&\delta_{i2}&\delta_{i7}&&&&&&\\
      \delta_{i8}&\delta_{i9}&\delta_{i3}&&&&&&\\
      &&&&\delta_{i10}&&&&\\
      &&&\delta_{i11}&&&&&\\
      &&&&&&\delta_{i12}&&\\
      &&&&&\delta_{i13}&&&\\
      &&&&&&&&\delta_{i14}\\
      &&&&&&&\delta_{i15}&\\
    \end{array}
  \right)
  \,.
\end{eqnarray}
This is a complete basis for a residual $U_{c+V}(1)\otimes U_{c+V}(1)$ symmetry
(see Eq.~(\ref{eq:CFLsymmetries})). The basis is
defined by
\begin{eqnarray*}
  \{(r,u),(g,d),(b,s),(r,d),(g,u),(r,s),(b,u),(g,s),(b,d)\},
\end{eqnarray*}
with $r$, $g$, $b$ denoting the color and $u$, $d$, $s$ the flavor of the quarks.

\end{document}